\documentclass[aps,prl,twocolumn,groupedaddress,showpacs]{revtex4}
\usepackage{graphicx}
\usepackage{dcolumn}
\usepackage{bm}
\begin{document}

\title{Demonstration of a moving guide based atom interferometer for rotation sensing}

\author{Saijun Wu, Edward Su, Mara Prentiss}
\affiliation{Department of Physics, Harvard University, Cambridge,
MA, 02138}

\affiliation {School of Engineering and Applied Science, Harvard
University, Cambridge, MA, 02138}
\date{\today}

\begin{abstract}
We demonstrate area-enclosing atom interferometry based on a moving
guide. Light pulses along the free propagation direction of a
magnetic guide are applied to split and recombine the confined
atomic matter-wave, while the atoms are translated back and forth
along a second direction in 50~ms. The interferometer is estimated
to resolve ten times the earth rotation rate per interferometry
cycle. We demonstrate a ``folded figure 8'' interfering
configuration for creating a compact, large-area atom gyroscope with
multiple-turn interfering paths.
\end{abstract}


\maketitle
\section{}

In a Sagnac interferometry based gyroscope, a rotation with an
angular velocity $\Omega$ induces a shift of the interference fringe
by the Sagnac phase $\varphi_{Sagnac}=\frac{2 m  \Omega A}{\hbar}$,
with $m$ the total mass of the interfering particle, $\hbar$ the
reduced Planck constant, and $A$ the area enclosed by the
interfering paths projected along the rotation axis. Due to the
large total mass, the Sagnac phase for atoms with the same area $A$
is typically $10^{10}$ times larger than it is for visible photons,
which makes the atomic matter-wave gyroscope a promising
candidate~\cite{Clauser88,Gustavson97} to replace laser
gyroscopes~\cite{RinglaserReview} to deliver unprecedented
rotational sensitivities, with applications expected in
long-distance inertial navigation, in geophysics research, and
potentially in testing fundamental theories such as to measure
geodesic and frame-dragging effects~\cite{Clauser88}.

Existing atom gyroscopes use diffractive optics to manipulate atomic
beams. In a typical 3-grating setup~\cite{Gustavson97}, the area
enclosed by the interfering paths can be expressed as
$A=\frac{1}{2}v_s\times v_{beam} T^2$, with $v_s$ the recoil
velocity associated with the grating diffraction, $v_{beam}$ the
atomic beam velocity, and $T$ the time during which the atoms are
successively interrogated by the three gratings. For example, the
state-of-the-art atom gyroscope developed in the Kasevich
group~\cite{Gustavson2000} uses cesium atomic beams with an
enclosed-area of $A\sim 22$~mm$^2$, based on an apparatus that is
$L=v_{beam}T\sim 2$~m long. Further increasing the area of atomic
gyroscopes would improve their rotational sensitivity. However, atom
gyroscopes need to be as compact as possible to suppress phase
shifts due to stray fields as well as those due to mechanical
vibrations, and a longer apparatus would limit the applications of
the gyroscope. Recent interferometry techniques have been developed
with slow atoms in parabolic trajectories~\cite{SixAxis06}. However,
a significant improvement of the sensing area in fountain geometry
requires a tall atomic fountain and thus a large apparatus.

An obvious solution to fulfill the contradictory requirements of
being compact and having a large sensing area is to let the two
interfering paths circulate around a small physical area multiple
times. To suppress any phase shift due to static perturbations, the
paths may be chosen to be reciprocal, with one path following the
time-reversal path of the other. A multiple-turn reciprocal
interference configuration may be achievable using atomic guiding
potentials. Recent research efforts have realized atomic wave-guide
potentials in closed loops~\cite{rings}. However, interferometry
with these devices has not been demonstrated. One of the technical
challenges of this approach stems from the dispersive coupling of
motion between the confining direction and the guide direction in
curved atomic wave-guides~\cite{betatron06}, usually large even in
the adiabatic limit. Recently the interference of a propagating
Bose-Einstein condensate was demonstrated to enclose an area
following a ``Y'' splitting scheme~\cite{chiplongcoherence};
however, it is unclear how to use the scheme to enclose a large area
in multiple-turn reciprocal geometry.
\begin{figure}
\centering
\includegraphics [width=3.5 in,angle=0] {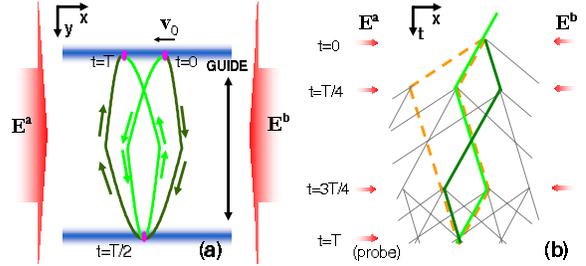}\\
\caption{(Color online) (a) Schematic of a moving guide
interferometer. The two blue bars represent the position of the
moving guide at times $t=0, T$ (top) and $t=T/2$ (bottom), with the
center-of-mass position of a guided atomic wavepacket at the three
times marked. The optical fields ${\bf E}^a$ and ${\bf E}^b$ are
pulsed (see b). The atom follows the thick solid curve to interfere
at time $t=T$. (b) The x-t recoil diagram for the 4-pulse scheme. At
time $t=T$ the atomic fringe is probed via a Bragg-scattering from
${\bf E}^a$ to ${\bf E}^b$. The ``figure 8'' interfering loop marked
with thick solid lines corresponds to that in (a). Another loop
marked with dashed lines is referred to as a ``trapezoid'' loop.
}\label{figscheme}
\end{figure}

Instead of transporting atoms with curved wave-guides, we
demonstrate an area-enclosing guided atom interferometer in a
``folded figure 8'' configuration based on a straight
guide~\cite{WangMichelson05} that moves. In particular, a
two-dimensional interfering path of guided atoms is created by a
pulsed optical standing wave field along $\bf e_x$ combined with a
guiding potential oscillating along $\bf e_y$
(Fig.~\ref{figscheme}a). A 4-pulse de-Broglie wave interferometry
scheme (Fig.~\ref{figscheme}b)~\cite{TLCahn97, mythesis} is applied
to create an effectively-reciprocal interfering loop that encloses
an area up to $0.2$~mm$^2$ with stable phase readouts. We argue that
the scheme demonstrated in this work can be extended to a practical
large-area guided atom gyroscope with multiple-turn reciprocal
paths.

We consider the schematic setup in Fig.~\ref{figscheme}. Atoms are
confined in a guiding potential oriented to $\bf e_x$, the direction
along which the guiding potential is invariant. An optical standing
wave field composed of traveling light ${\bf{E}}^a$ and ${\bf{E}}^b$
is aligned parallel to $\bf e_x$~\cite{WangMichelson05}. Standing
wave pulses split and recombine the atom wavepackets along $\bf e_x$
by transferring photon recoil momentum to atoms. When atoms are
transported along the direction $\bf e_y$, the interfering paths
enclose the area in the $\bf x-y$ plane. In particular, as
represented by the recoil diagrams shown by the thick solid lines in
Fig.~\ref{figscheme}a, b, a nearly reciprocal interfering loop can
be created by setting the guide velocity $v_y \propto \sin
(\frac{{2\pi t}}{T})$ while the atom wavepackets are split at time
$t=0$ and redirected at time $t=T/4$, $3T/4$. Pairs of wavepackets
meet at time $t=T$, creating a ``folded figure 8'' loop in the x-y
diagram and a ``figure 8'' loop in the x-t diagram. The area
enclosed by the loop is $A = \frac{L}{\pi }\delta v_x T$ , where
$\delta v_x$ is the wavepacket splitting velocity due to photon
recoils and $L$ is the maximum translation distance along
${\bf{e}}_{\bf{y}}$. Since the atoms almost return to the initial
position at time $T$, the sequence may be repeated $N$ times; each
time the guide passes the half distance $L$ the two interfering
paths are deflected by an additional standing wave pulse, resulting
in multiple turn interfering paths. Precise reciprocity of the
interfering paths can be achieved with atoms starting with zero
velocity. With non-zero initial velocity $v_0$
(Fig.~\ref{figscheme}a), effective reciprocity is retained that
suppresses the interferometry phase shifts due to potential
variations at the length scale $l$ with $v_0 T << l$.

\begin{figure}
\centering
\includegraphics [width=1.6in,angle=270] {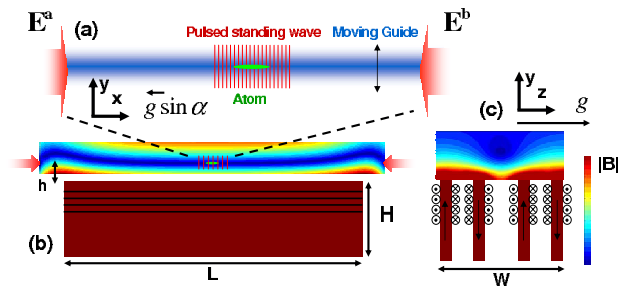}\\
\caption{Schematic of the moving guide interferometry setup in this
work. (a) is a zoom-in of (b) at the standing wave zone. (b) and (c)
show the arrangement of the permalloy foils and the current-carrying
wires. A density plot of the simulated magnetic guiding potential is
included in (b) and (c).}\label{fig4foil}
\end{figure}

In this demonstration, a 4-pulse de-Broglie wave interferometry
scheme (Fig.~\ref{figscheme}b)~\cite{gratingechoOrigin, TLCahn97} is
applied to address most of the atoms in our 25~$\mu K$ atomic
sample. We use short standing wave pulses (300~ns width) as thin
phase gratings, that diffract atoms to multiple diffraction orders
weighted by the amplitude $i^n J_{n}(\theta)$. Here $\theta$ is the
standing wave pulse area and $J_n $ is the $n^{th}$ order Bessel
function~\cite{TLCahn97}. Pairs of paths interfere and create an
atomic density grating at time $T$, which is probed by monitoring
the Bragg scattering of a probe light from ${\bf{E}}^a$ into
${\bf{E}}^b$ mode~\cite{mythesis}. The two requirements for any two
interfering paths to contribute to the Bragg backscattering of a
probe light (grating echo) are: 1) they meet at time $T$ to form a
loop, and 2) at time $T$ their relative velocities should be twice
the atomic recoil velocity $2 v_r\approx 11.8$~mm/sec. We classify
these loops according to the relative displacements between each
pair of paths. In addition to the ``figure 8'' loop in
Fig.~\ref{figscheme}b that corresponds to the ``folded figure 8''
loop in Fig.~\ref{figscheme}a, other interfering loops also
contribute, such as the ``trapezoid'' loop in Fig.~\ref{figscheme}b.
The final interferences are composed of the contributions from all
these loops. It can be shown that the contributions from the
``figure 8'' and the ``trapezoid'' loops are weighted by $J_2^2
(2\theta )$ and $J_1^2 (2\theta )$ respectively~\cite{mythesis},
while other loops contribute negligibly for standing wave pulse area
$\theta < 2$. The reciprocity of the ¡°figure 8¡± loop ensures a
zero differential phase shift between the two paths in the presence
of a linear potential along ${\bf e_x}$, and suppresses the
dephasing effects due to nonlinear potential variations. This is in
contrast to the ¡°trapezoid¡± loop that acquires a phase shift of
$\varphi (T) = \frac{3}{8}k a T^2 $  due to an acceleration force $m
a {\bf e_x}$ ($k$ the wave-vector of the light field) and dephases
quickly in a nonlinear potential. The reciprocity of the ``figure
8'' loop is confirmed experimentally by analyzing the readouts due
to the ``trapezoid'' loop and the ``figure 8'' loop.


In what follows, we briefly summarize our experimental apparatus,
detailed in~\cite{mythesis}. The layout of the experimental setup is
described in Fig.~\ref{fig4foil}. Four 200~mm $\times$ 100~mm
$\times$ 1.5~mm permalloy foils, separated by 6.35, 12.7 and
6.35~mm, are poled in alternating directions (Fig.~\ref{fig4foil}b,
c) to generate a quadruple field as the guiding potential. Close to
the center of the foils, the quadruple field can be approximately
described by ${\bf{B}} = B_1 (z{\bf{e}}_y + y{\bf{e}}_z ) + B_0
{\bf{e}}_{\bf{x}} $. A wave-guide operation distance of $h\sim
7$~mm~$<<L$~=~200~mm is chosen to minimize the edge effects. With
the wave-guide operating at a field gradient of $B_1$=70~G/cm, $B_0$
varies about 10~mG over a centimeter along $\bf e_x$, likely due to
the foil-surface inhomogeneities. Approximately $10^{7}$
laser-cooled $^{87}$Rb atoms in their ground state $F=1$ hyperfine
level are loaded into this magnetic guide resulting in a
cylindrically-shaped atom sample 1~cm long and 170~$\mu m$ wide. The
standing wave is formed by two counter-propagating laser beams ${\bf
E}^a $ and ${\bf E}^b $ (120~MHz to the blue side of the F=1 -
F$^\prime$=2 D2 transition) with diameters of 1.6~mm. A heterodyned
detection of the Bragg scattering is applied to retrieve both the
amplitude and the phase of the atomic fringe~\cite{TLCahn97}.
Precise alignment of the standing wave k-vector parallel to the
free-propagation direction of the guide is required to decouple the
wave-guide confinement from the interferometry phase
readouts~\cite{mythesis}. To ensure parallelism, the relative angle
between the standing wave and the wave-guide is minimized with two
rotation stages (with the rotation axis along $\bf e_y$ and $\bf
e_z$ in Fig.~\ref{fig4foil}) at a precision better than $0.2$~mrad
by minimizing the confinement-induced interferometry
dephasing~\cite{GuideTalbot05}.

\begin{figure}
\centering
\includegraphics [width=1.4 in,angle=270] {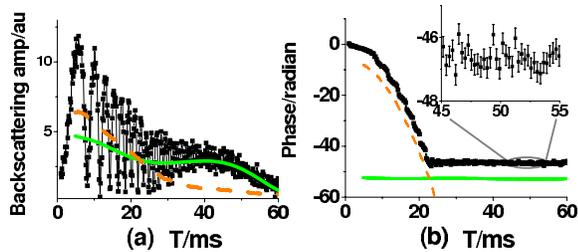}\\
\caption{(Color online) 4-pulse interferometry readout in a
stationary guide. The 4-pulse data (black) is decomposed into the
solid  and the dash curves, that corresponds to the contributions
from the ``figure 8'' and ``trapezoid'' loops in
Fig.~\ref{figscheme} respectively. (a) Interferometry amplitude (b)
Interferometry phase with an inset plot around 50~ms. The phases in
(b) have been unwrapped. The error bars give the phase noise due to
mirror vibrations.}\label{figst}
\end{figure}

To illustrate the interference and decay of different interfering
loops in Fig.~\ref{figscheme}b, we first discuss interferometry
output with atoms in the stationary guide that are shown in
Fig.~\ref{figst}. Given the 300~ns standing wave pulse duration, we
set the pulse area to be $\theta=1.3$ for the second and the third
pulse by adjusting the intensity of the pulses. In repeated
experiments, the amplitude and phase of the Bragg scattering signals
were recorded with different total interrogation time $T$, which is
incremented from $T=0.9945$~ms to $T=60.1314$~ms in steps of
132.6~$\mu$s. The oscillatory amplitude/phase readout in
Fig.~\ref{figst} is due to the beat between the ``figure 8'' loop
(thick solid line) and the ``trapezoid'' loop ( thick dashed line)
in Fig.~\ref{figscheme}b. With a numerical routine to separate the
fast oscillating parts from the slowly varying parts, the
oscillatory interferometry amplitude and phases are decomposed to
the solid and dashed curves in Fig.~\ref{figst}, which represent the
contributions from the ``figure 8'' and the ``trapezoid'' loops in
Fig.~\ref{figscheme}b respectively. The acceleration force is due to
a small gravity component along the direction $\bf e_x$ with $a =
g\sin \alpha $ (In Fig.~\ref{fig4foil} the gravity is along
${\bf{e}}_{\bf{z}}  + \alpha {\bf{e}}_{\bf{x}}$ and $\alpha \sim
3$~mrad). We extract the acceleration constant $a =
31.0(5)$~mm/$\sec ^2$ from the dashed curve in Fig.~\ref{figst}b,
which is found to agree with those from independent measurements
using 3-pulse interferometers~\cite{TLCahn97}. The solid curve in
Fig.~\ref{figst}a has a smaller initial amplitude but a slower
decay, so that the two curves cross at $T\sim22$~ms.
Correspondingly, the phase readout in Fig.~\ref{figst} approximately
follows the parabolic curve for $T<22$~ms but is locked to a
constant value for $T>22$~ms. The slow dephasing of the ``figure 8''
loop is attributed to the approximate reciprocity between the two
interfering paths (Fig.~\ref{figscheme}b). The results in
Fig.~\ref{figst} demonstrate that in the presence of multi-loop
interference in this experiment, an independent observation of the
``figure 8'' loop is nevertheless possible for $T\sim 50$~ms due to
its relatively long coherence time.

\begin{figure}
\centering
\includegraphics [width=3.4 in,angle=0] {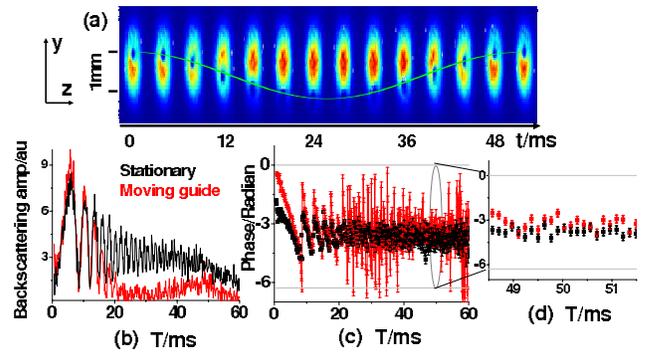}\\
\caption{Top: Absorption images of the guided atomic sample
following the moving guide in 50~ms. The graphs are expanded along
the $\bf e_y$ direction for clarity. A ``cosine'' curve is added as
a guide for the readers' eyes. Bottom: Interferometer signals for
the case of stationary (black) and moving (red) guides. In (c), (d)
the error bars indicate the standard phase deviation due to mirror
vibrations. Notice the scale difference between Fig.~\ref{figst}b
and (c) here. }\label{figapp}
\end{figure}

We now discuss the moving guide interferometer. To translate the
atomic sample along $\bf e_y$, we pulse a 50ms-period sinusoidal
current through the coils that magnetize the inner two foils of the
4-foil (Fig.~\ref{fig4foil}c). The resulting motion of the guided
atomic sample in the $\bf y-z$ plane is monitored with the
absorption images probed by $\bf E^b$ (Fig.~\ref{fig4foil}a, b). For
the absorption image, the frequency of $E^b$ is tuned to resonance
before the laser is coupled into a fiber. The fiber coupling ensures
that the spatial mode of $E^b$ is consistent for both the absorption
image and the standing wave formation. A sequence of absorption
images taken in repeated experiments is presented in
Fig.~\ref{figapp}a. The motion of atoms follows a ``cosine'' trace
fairly well with an amplitude of $0.97(5)$~mm.

To realize area-enclosing interferometry, we repeat the 4-pulse
interferometry measurements with the same current settings that
induced the atomic motion shown in Fig.~\ref{figapp}a. Typical
interferometry amplitude and phase readouts are plotted in
Fig.~\ref{figapp} in red, together with comparison data from another
stationary guide experiment plotted in black. Here a standing wave
pulse area of $\theta=1.5$ is chosen to maximize the interference
due to the ``figure 8'' loop. A consequence is the more constant
phase readout for the black plots in Fig.~\ref{figapp}c, compared
with those shown in Fig.~\ref{figst}b. In contrast to the stationary
guide case, the amplitude of the moving guide interferometer signal
(red plots) almost vanishes in the time window of $T$ from 25~ms to
35~ms because the atoms are at the very outer edge of the optical
fields when the probing pulse is fired, resulting in vanishing
backscattering signals and thus random phase readouts
(Fig.~\ref{figapp}c). For $T\sim 50$~ms, the atomic sample returns
back to within the $\sim 1.6$~mm diameter of the optical fields. The
backscattered signal revives and the phase readout matches those in
the stationary guide case (Fig.~\ref{figapp}d). Similar to
Fig.~\ref{figst}b, the constant and repeatable phase readouts in
Fig.~\ref{figapp}d are due to the approximate reciprocity of the
``figure 8'' loop, which in the spatial-domain becomes the ``folded
figure 8'' loop in Fig.~\ref{figscheme}a. For $T=50$~ms, the
interferometer encloses an area of 0.18~mm$^{2}$~\cite{exp:foot1}. A
rotation rate of 1~mrad/sec should induce a Sagnac phase of $\pi$
that shifts the red plot in Fig.~\ref{figapp} d.

The small phase offset observed in Fig.~\ref{figapp}d between the
readouts from a stationary guide and the moving guide is repeatable
before any realignments. The offset is due to the guiding potential
variation along $\bf e_x$. In this experiment, the moving guide
potential does not precisely follow a sinusoidal trace, and
collective atomic oscillations in the moving guide are expected
since the wave-guide confinement is relatively loose. The
non-reciprocity leads to imperfect cancelation of the phase shifts
before and after $t=T/2$ (Fig.~\ref{figscheme}) that results in the
phase offset in Fig.~\ref{figapp}d. The imperfection is not
precisely repeatable in each experimental trial, resulting in a
standard deviation of the phase readout $\sim
0.4$~rad~\cite{exp:foot2}. A smoother and more precisely controlled
guiding potential would help to suppress the phase error as well as
the dephasing effects that have limited the interrogation time of
50~ms in this experiment. By reducing the wavepacket separations
using a different interferometry scheme, we have achieved up to
1~$\sec$ coherence time with guided atoms~\cite{longcoherence}.

The $\sim 0.2$~mm$^2$ enclosed area is much smaller than
in~\cite{Gustavson2000}. However, orders of magnitude improvements
in the sensing area are expected with larger guide translation
distances, a more efficient beamsplitting scheme, and longer
interrogation time~\cite{mythesis}. To increase the wave-guide
translation distance, a multi-wire 1D conveyer belt on an atom
chip~\cite{WangMichelson05} may be constructed. The localized atomic
sample in the wave-guide should facilitate a high-efficiency
multiple-recoil beamsplitting~\cite{doublepulse05} where the light
intensity control is important. We consider a $^{87}$Rb atomic
sample at sub-recoil temperatures so that multi-photon beam
splitting techniques can be applied giving efficient $\pm 6\hbar k$
momentum splittings~\cite{doublepulse05}. If the wave-guide were
oscillated 5 cm back and forth 5 times accompanied by 10
appropriately timed Bragg pulses, the interferometer on a
centimeter-scale device would enclose an area greater than
$\sim$1000~mm$^2$ in a second. If we consider a shot-noise limited
phase resolution, the resulting on-chip atom gyroscope would have a
rotational sensitivity of $1\times 10^{ - 9}$~rad/$\sec \sqrt {\rm
Hz}$ comparable to~\cite{Gustavson2000}, even with only $10^4$ atoms
per experimental cycle. Parallel operation of multiple guided atom
gyroscopes may further boost the measurement bandwidth and
sensitivity.

In conclusion, we have demonstrated a moving-guide based atom
interferometer that encloses an area of $0.2$~mm$^2$ and has stable
phase readout for rotation sensing. We have demonstrated a ``folded
figure 8'' interferometry configuration whose reciprocity partly
suppresses matterwave dephasing due to guiding potential variations.
The ``folded figure 8'' configuration should be sensitive to
rotation but extremely insensitive to linear acceleration. The
scheme may be extended to enable sensitive rotation measurement with
a multiple-turn atomic matterwave gyroscope in a compact device.

This work also demonstrates coherent transportation of matter-waves
using magnetic-dipole forces. The moving guide may allow a light
pulse atom interferometer to interact with a distant object where
the optical path is restricted, such as to sense the light shift of
a micro-cavity or a surface potential.

\begin{acknowledgments}
This work is supported by MURI and DARPA from DOD, NSF, ONR and U.S.
Department of the Army, Agreement Number W911NF-04-1-0032, and by
the Charles Stark Draper Laboratory. We thank the referees for their
suggestions on the earlier version of this article.
\end{acknowledgments}



\begin{thebibliography}{99}
\bibitem{exp:foot1}
Area up to $0.2$~mm$^2$ was enclosed with $T=54$~ms.
\end{thebibliography}
\end{document}